%% file: main_final.tex
\documentclass[12pt]{article}

\usepackage{scicite}
\usepackage{times}
\usepackage{graphicx}  
\usepackage{dcolumn}   
\usepackage{bm}        
\usepackage{amssymb}   
\usepackage{amsmath}   
\usepackage{amsthm}    
\usepackage{xcolor}    
\usepackage{tikz}      
\usepackage{ifthen}    
\usepackage{multirow}  
\usepackage{tabu}      
\usepackage{subfigure}
\usepackage{xspace}
\usepackage{comment}
\usepackage[unicode=true,
  linktocpage,
  linkbordercolor={0.5 0.5 1},
  citebordercolor={0.5 1 0.5},
  linkcolor=blue]{hyperref}
\newcommand{\urusi}{URu$_2$Si$_2$\xspace}
\newcommand{\THO}{$\mathrm{T}_\mathrm{HO}$\xspace}

\newcounter{para}

\newenvironment{sciabstract}{%
	\begin{quote} \bf}
	{\end{quote}}

\begin{document}

\title{One-Component Order Parameter in URu$_2$Si$_2$ Uncovered by Resonant Ultrasound Spectroscopy and Machine Learning}

\author{Sayak Ghosh,$^{1\dagger}$ Michael Matty,$^{1\dagger}$ Ryan Baumbach,$^2$ Eric D. Bauer,$^3$\\ K. A. Modic,$^4$ Arkady Shekhter,$^2$ J. A. Mydosh,$^5$\\ Eun-Ah Kim$^1$ and B. J. Ramshaw$^{1\ast}$\\
\\
\normalsize{$^{1}$Laboratory of Atomic and Solid State Physics,}\\ \normalsize{Cornell University, Ithaca, NY, 14853}
\\
\normalsize{$^{2}$National High Magnetic Field Laboratory,}\\ \normalsize{Florida State University, Tallahassee, FL 32310, USA}
\\
\normalsize{$^3$Los Alamos National Laboratory, Los Alamos, NM, USA 87545.}
\\
\normalsize{$^4$Max-Planck-Institute for Chemical Physics of Solids, Dresden, Germany 01187.}
\\
\normalsize{$^5$Kamerlingh Onnes Laboratory and Institute Lorentz,} \\ \normalsize{Leiden University, 2300 RA Leiden, The Netherlands.}
\\
\normalsize{$\dagger$ Authors contributed equally to this work.}
\\
\normalsize{$^\ast$To whom correspondence should be addressed: bradramshaw@cornell.edu.}
}
\date{}

\baselineskip24pt
\maketitle

\begin{sciabstract}
The unusual correlated state that emerges in URu$_2$Si$_2$ below \THO = 17.5 K is known as “hidden order” because even basic characteristics of the order parameter, such as its dimensionality (whether it has one component or two), are ``hidden". We use resonant ultrasound spectroscopy to measure the symmetry-resolved elastic anomalies across \THO. We observe no anomalies in the shear elastic moduli, providing strong thermodynamic evidence for a one-component order parameter. We develop a machine learning framework that reaches this conclusion directly from the raw data, even in a crystal that is too small for traditional resonant ultrasound. Our result rules out a broad class of theories of hidden order based on two-component order parameters, and constrains the nature of the fluctuations from which unconventional superconductivity emerges at lower temperature. Our machine learning framework is a powerful new tool for classifying the ubiquitous competing orders in correlated electron systems.
\end{sciabstract}

\section*{Introduction}

Phase transitions mark the boundary between different states of matter, such as liquid to solid, or paramagnet to ferromagnet. At the phase transition the system lowers its symmetry: translationally invariant liquids become crystalline solids; paramagnetic spins align to break time reversal and rotation symmetry in a magnet. The conventional description of a second-order phase transition---Landau theory---requires knowledge of which symmetries are broken in the low-temperature phase to construct an order parameter (OP). Several possibilities have been put forth for the symmetry of the OP in the hidden order (HO) state of \urusi (\autoref{tab:order}), but most of these rely on specific microscopic mechanisms that are difficult to verify experimentally \cite{mydosh:rmp2011a,mydosh:pm2014a}.

The purpose of this paper is to use resonant ultrasound spectroscopy (RUS) to place strict thermodynamic constraints---independent of microscopic mechanism---on the OP symmetry in \urusi. While RUS is a powerful technique---capable of constraining or identifying the symmetries broken at a phase transition \cite{RamshawPNAS}---it has one significant drawback: a single missing resonance renders an entire spectrum unusable. This is because traditional RUS data analysis relies on solving the elastic wave equation and mapping the computed resonances one-to-one with measured resonances---a single missing resonance invalidates this mapping. Here we develop a new machine-learning based approach. We take advantage of the fact that neural networks can be trained to recognize features in complex data sets and classify the state of matter that produce such data \cite{carrasquilla:np2017a,Ouyang:2018,Bohrdt:2018,Rem:2018,Zhang:2018,Yamaji:2019}. We validate this approach by analyzing an RUS data set that we are confident can also be analyzed using traditional methods (data from a large single-crystal of \urusi with a well-defined geometry). We then analyze data from a higher quality \urusi sample that has an ill-defined geometry---a task that is impossible for the traditional analysis method, but which is easily performed by our neural network.

While the broken symmetries of HO are unknown, most theories assume some form of `multipolar order', whereby localized $5f$ electrons on the uranium site occupy orbitals that order below \THO$= 17.5$~K. However, direct experimental evidence for localized $5f$ electrons---such as crystalline electric field level splitting---does not exist \cite{mydosh:rmp2011a}, leaving room for theories of HO based on itinerant $5f$ electrons. Many possible OPs remain in contention, but, whether itinerant or localized, all theories of HO can be classified based on the dimensionality of their point group representation: one-component \cite{harima:jpsj2010a,ohkawa:jpcm1999a,santini:prl1994a,kiss:ap2004a,haule:np2009a,kusunose:jpsj2011a,cricchio:prl2009a,Kung:2015,Kung:2016,kambe:pr2018a} or two-component \cite{thalmeier:pr2011a,hoshino:jpsj2013a,rau:pr2012a,tonegawa:prl2012a,fujimoto:prl2011a,ikeda:np2012,riggs:nc2015a} (see \autoref{tab:order} and footnote \cite{note1}). Theories of two-component OPs are motivated largely by the experiments of Okazaki et al. \cite{okazaki:s2011a} and Tonegawa et al. \cite{tonegawa2014direct}, which detect a small $C_4$ symmetry breaking at \THO. More recent X-ray experiments have cast doubt on these results \cite{Choi:2018}, leaving even the dimensionality of the OP in \urusi an open question.

Determining OP dimensionality is more than an exercise in accounting: the two-component nature of loop currents allows for dynamics that have been suggested to explain the pseudogap in high-T$_{\mathrm{c}}$ cuprates \cite{Aji:2010}; and the proposed two-component $p_x+i p_y$ superconducting state of Sr$_2$RuO$_4$ has a unique topological structure that can support Majorana fermions \cite{Rice:1995,Read:2000}. Establishing the dimensionality of the HO state not only allows us to rigorously exclude a large number of possible OPs, it also provides a starting place for understanding the unusual superconductivity that emerges at lower temperature in \urusi.

\begin{table*}
        \renewcommand{\arraystretch}{1.00}
				\footnotesize
        \begin{tabular}{||c|c|c||}
        \hline
			Dimensionality & Symmetry & Reference \\ 
			\hline
			\multirow{7}{3.5em}{One-component} & A$_{1g}$ & Harrison and Jaime\cite{Harrison:2019} \\
			\cline{2-3}
			& A$_{1u}$ & Kambe \textit{et al.}\cite{kambe:pr2018a} \\
			\cline{2-3}
			& A$_{2g}$ & Haule and Kotliar\cite{haule:np2009a}, Kusunose and Harima\cite{kusunose:jpsj2011a},  Kung \textit{et al.}\cite{Kung:2015}  \\ 
			\cline{2-3}
			& A$_{2u}$ & Cricchio \textit{et al.}\cite{cricchio:prl2009a} \\
			\cline{2-3}
			& B$_{1g}$ & Ohkawa and Shimizu\cite{ohkawa:jpcm1999a}, Santini and Amoretti\cite{santini:prl1994a} \\
			\cline{2-3}
			& B$_{1u}$ & Kiss and Fazekas\cite{kiss:ap2004a} \\
			\cline{2-3}
			& B$_{2g}$ & Ohkawa and Shimizu\cite{ohkawa:jpcm1999a}, Santini and Amoretti\cite{santini:prl1994a}, Harima \textit{et al.}\cite{harima:jpsj2010a} \\
			\cline{2-3}
			& B$_{2u}$ & Kiss and Fazekas\cite{kiss:ap2004a} \\
			\hline 
			\multirow{2}{3.5em}{Two-component} &E$_{g}$ & Thalmeier and Takimoto\cite{thalmeier:pr2011a}, Tonegawa \textit{et al.}\cite{tonegawa:prl2012a}, Rau and Kee\cite{rau:pr2012a}, Riggs \textit{et al.}\cite{riggs:nc2015a} \\
			\cline{2-3}
			& E$_{u}$ &  Hoshino \textit{et al.}\cite{hoshino:jpsj2013a}, Ikeda \textit{et al.}\cite{ikeda:np2012}, Riggs \textit{et al.}\cite{riggs:nc2015a} \\
				\cline{2-3}
			& E$_{3/2,g}$ & Chandra \textit{et al.}\cite{Chandra2013} \\
        \hline	
		\end{tabular}
		\caption{\label{tab:table1} Proposed order parameters of the HO state in \urusi, classified by their dimensionality and their point-group representation. Note that designations such as ``hexadecapole'' order are only applicable in free space---crystalline electric fields  break these large multipoles into the representations listed in this table.}
		\label{tab:order}
	\end{table*}

\section*{Experiment}
\subsection*{Resonant Ultrasound Spectroscopy of \urusi}

RUS measures the mechanical resonance frequencies of a single-crystal specimen---analogous to the harmonics of a guitar string but in three dimensions (see Fig. 1a). A subset of this spectrum for a 3 mm x 2.8 mm x 2.6 mm crystal of \urusi (sample S1) is shown in Fig. 1b, with each peak occurring at a unique eigenfrequency of the elastic wave equation (see supplementary information). Encoded within these resonances is information about the sample's dimensions and density, which are known, and the six elastic moduli, which are unknown. As electrons and phonons are coupled strongly in metals, the temperature dependence of the elastic moduli reveals fluctuations and instabilities in the electronic subsystem. In particular, elastic moduli are sensitive to symmetry breaking at electronic phase transitions \cite{Shekhter:2013,RamshawPNAS}. The difficulty lies in converting the temperature dependence of the resonance spectrum into the temperature dependence of the elastic moduli. The traditional analysis involves solving the 3D elastic wave equation and adjusting the elastic moduli to match the experimental resonance spectrum. However, if even a single resonance is missing from the spectrum (e.g. due to weak coupling of a particular mode to the transducers), then this analysis scheme breaks down (see Ramshaw \textit{et al.}\cite{RamshawPNAS} for further discussion of this problem).

Fig. 1c shows the temperature dependence of seven representative elastic resonances through \THO (29 resonances were measured in total). Note that while some resonances show a step-like discontinuity or `jump' at \THO, others do not. This jump is present in the elastic moduli for all second-order phase transitions \cite{LuthiPRB,Shekhter:2013,RamshawPNAS}, but has never before been observed in \urusi due to insufficient experimental resolution \cite{Fukase:1987,Bullock:1990,Wolf:1994,Yanagisawa:2012,Yanagisawa:2018}. Traditional RUS produces spectra at each temperature, such as the one shown in Fig. 1b, by sweeping the entire frequency range using a lock-in amplifier. The resonance frequencies are then extracted by fitting Lorentzians to each peak \cite{Shekhter:2013}. We have developed a new approach whereby the entire spectrum is swept only once---to identify the resonances---and then each resonance is tracked as a function of temperature with high-precision using a phase-locked loop. This increases the density of data points per unit temperature by roughly a factor of 1000, and increases the signal to noise by a factor of 30 (see methods).

\begin{figure*}
	\centering
	\includegraphics[width=.99\linewidth]{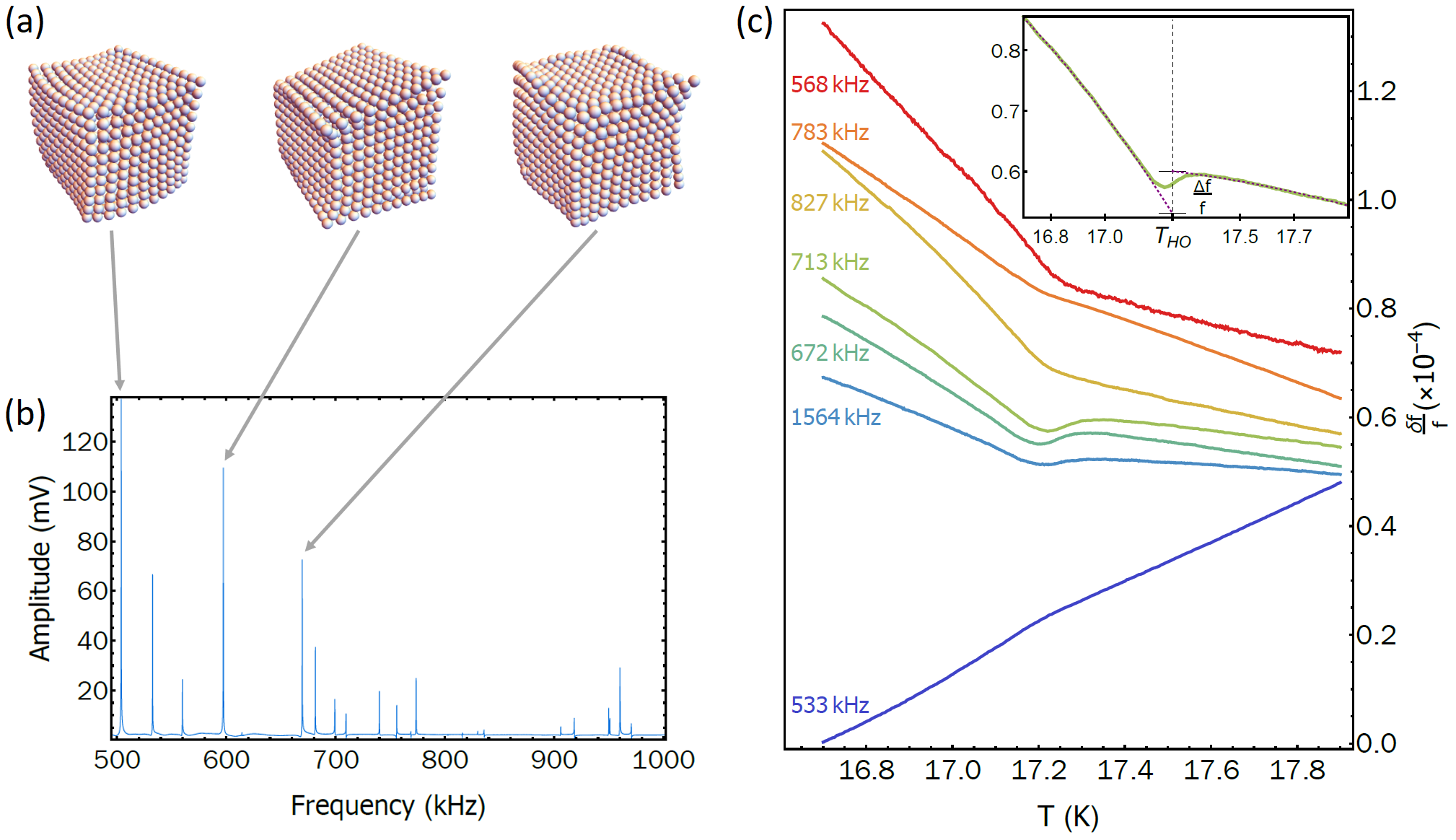}
	\caption{ \textbf{Resonant ultrasound spectroscopy across T$_{\boldsymbol{\mathrm{HO}}}$ in \urusi.} (a) Schematic resonance eigenmodes obtained as a solution to the 3D elastic wave equation. Each mode contains a unique proportion of the five irreducible strains (see Fig. 2a). (b) Room temperature ultrasonic spectrum of sample S1, shown between 500 kHz and 1 MHz. (c) Temperature evolution of seven characteristic resonances, out of 29 total measured resonances, near the HO transition---plots are shifted vertically for clarity. Three resonances (672 kHz, 713 kHz and 1564 kHz) show jumps at \THO (inset illustrates what is meant by the jump), while the others do not, signifying contributions from different symmetry channels.}
	\label{fig:resonances}
\end{figure*}

The complex strain fields produced at each resonance frequency (Fig. 1a) can be broken down locally into irreducible representations of strain ($\epsilon_k$). Each irreducible strain then couples to an OP $\eta$ of a particular symmetry in a straightforward manner \cite{LuthiPRB}. In this way, analysis of the temperature dependence of the resonance frequencies can identify or constrain the OP symmetry. In a tetragonal crystal, such as \urusi, elastic strain breaks into five irreducible representations (Fig. 2): two compressive strains transforming as the identity $A_{1g}$ representation, and three shear strains transforming as the $B_{1g}$, $B_{2g}$ and $E_{g}$ representations. Allowed terms in the free energy $\mathcal{F}$ are products of strains and OPs that transform as the $A_{1g}$ representation. As HO is thought to break at least translational symmetry, the lowest-order terms allowed by both one-component and two-component OPs are linear in the $A_{1g}$ strains and quadratic in OP:$~\mathcal{F} = \epsilon_{A_{1g}}\!\cdot \eta^2$ (see footnote \cite{note2}). Quadratic-in-order-parameter, linear-in-strain coupling produces a discontinuity in the associated elastic modulus at the phase transition: this jump is related to discontinuities in the specific heat and other thermodynamic quantities through Ehrenfest relations \cite{Shekhter:2013,Modic:2018b}. For OPs with one-component representations (any of the $A_i$ or $B_i$ representations of $D_{4h}$), only the elastic moduli corresponding to $A_{1g}$ compressional strains couple in this manner. In contrast, shear strains couple as $\mathcal{F} = \epsilon_{k}^2\!\cdot \eta^2$ and show at most a change in slope at \THO \cite{RamshawPNAS}. Thus $c_{33}$, $c_{23}$ and $(c_{11}+c_{12})/2$ may exhibit jumps at phase transitions corresponding to one-component OPs, while $(c_{11}-c_{12})/2$, $c_{66}$, and $c_{44}$ cannot. 

\begin{figure*}
	\centering
	\includegraphics[width=.99\linewidth]{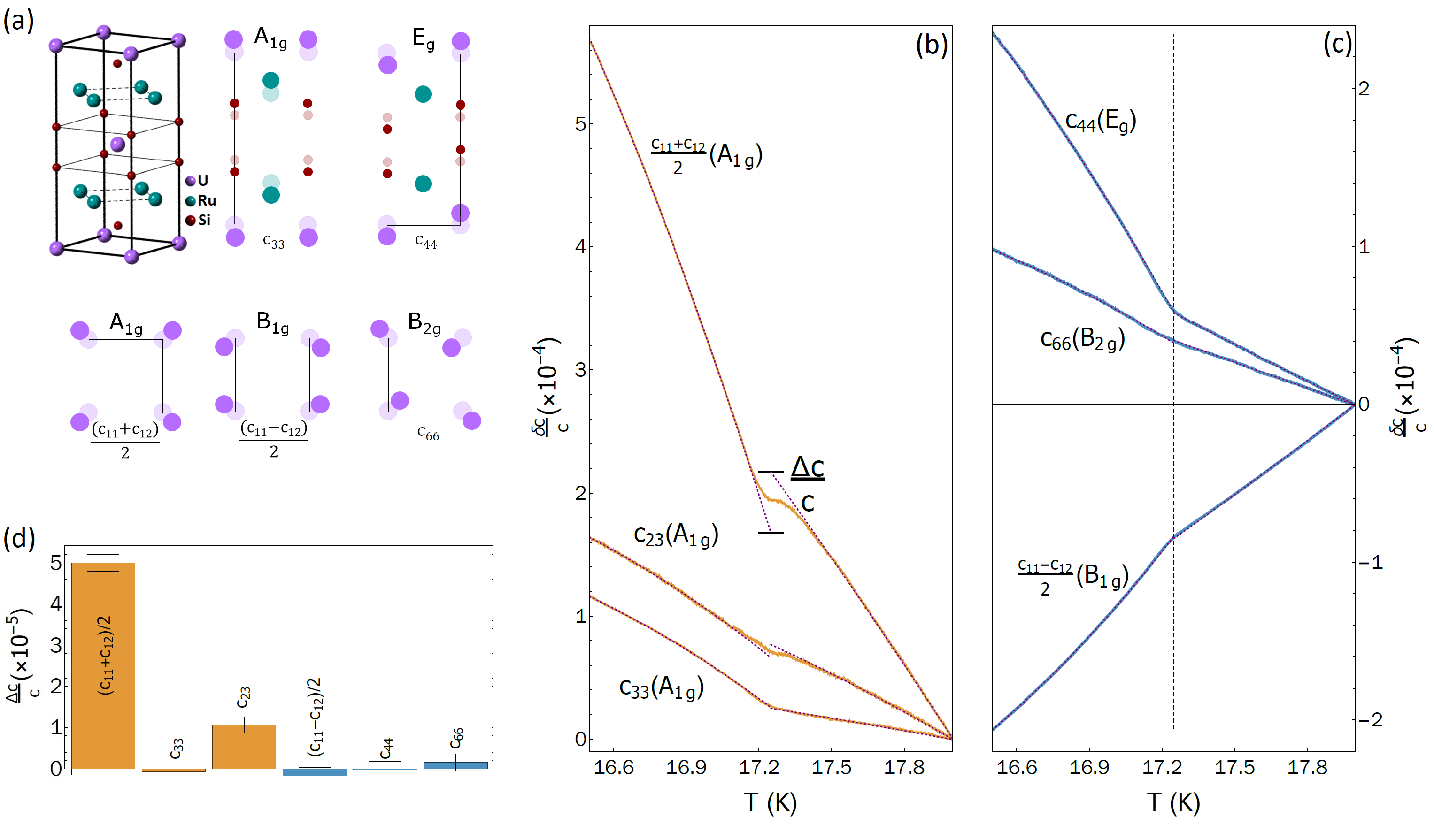}
	\caption{\textbf{Traditional extraction of symmetry information from elastic moduli.} (a) The tetragonal crystal structure of $\mathrm{URu_{2}Si_{2}}$ and its five irreducible representations of strain, along with the associated moduli. Each resonance shown in Fig. 1a can be decomposed into this basis set of strains, modulated in phase at long wavelengths throughout the crystal. $c_{23}$ characterizes the direct coupling between the two $A_{1g}$ strains. (b) Compressional ($A_{1g}$, shown in orange) and (c) shear ($B_{1g}$, $B_{2g}$, and $E_{g}$, shown in blue) elastic moduli, with dashed guides to the eye showing the temperature dependence extrapolated from below and above \THO. The absolute values (in GPa) of the moduli at $\sim$20~K were determined to be $(c_{11}+c_{12})/2 = 218.0$, $c_{33}=307.4$ , $c_{23}=112.8$, $(c_{11}-c_{12})/2 =65.2 $, $c_{66}=140.6 $ and $c_{44}=101.8$. (d) The magnitude of the jumps at \THO with their experimental uncertainties. A large jump occurs in  $(c_{11}+c_{12})/2$ at \THO, along with a small jump in $c_{23}$. The shear moduli, on the other hand, show only a change in slope at \THO---this constrains the OP of the HO state to transform as a one-component representation.}
	\label{fig:moduli}
\end{figure*}

Two-component OPs (of the $E_i$ representations), on the other hand, have bilinear forms that can couple with two of the shear strains to first order. A two-component OP, $\vec{\eta} = \left\{\eta_x,\eta_y\right\}$, has the bilinears $\eta_x^2 + \eta_y^2$, $\eta_x^2 - \eta_y^2$, and $\eta_x \eta_y$; of the $A_{1g}$, $B_{1g}$, and $B_{2g}$ representations, respectively. In addition to the standard $\epsilon_{A_{1g}}\!\cdot \eta_x^2 + \eta_y^2$ terms, the free energy now contains the terms $\epsilon_{B_{1g}}\!\cdot\left( \eta_x^2 - \eta_y^2\right)$ and $\epsilon_{B_{2g}}\!\cdot \eta_x\eta_y$. A second order phase transition characterized by a two-component OP therefore exhibits discontinuities in the $B_{1g}$ and $B_{2g}$ shear elastic moduli ( $(c_{11}-c_{12})/2$ and $c_{66}$, respectively), in addition to jumps in the compressional $A_{1g}$ moduli (see SI for a discussion of the $E_{3/2,g}$ representation, pertaining to ``hastatic'' order). 

We first perform a traditional RUS analysis, extracting the temperature dependence of the six elastic moduli (Fig. 2b and c) from 29 measured resonances by solving the elastic wave equation and fitting the spectrum using a genetic algorithm (see SI of Ramshaw \textit{et al.}\cite{RamshawPNAS} for details). The evolution of the elastic moduli across \THO shows jumps in two of the $A_{1g}$ elastic moduli, whereas the $B_{1g}$ and $B_{2g}$ shear moduli show only a break in slope at \THO to within our experimental uncertainty (Fig. 2d). Jumps in the shear moduli would be expected for any order parameter of the two-component $E_i$ representations \cite{thalmeier:pr2011a,hoshino:jpsj2013a,rau:pr2012a,tonegawa:prl2012a,fujimoto:prl2011a,ikeda:np2012,riggs:nc2015a}---the fact that we do not resolve any shear jumps constrains the OP of the HO phase to belong to a one-component representation of $D_{4h}$. The fact that we do not resolve a jump in $c_{33}$ is consistent with the magnitudes of the jumps in $(c_{11}+c_{12})/2$ and $c_{23}$---see SI for details.
 

In principle this traditional analysis is sufficient to determine the order-parameter dimensionality in \urusi. The process of solving for the elastic moduli, however, incorporates systematic errors arising from sample alignment, parallelism, dimensional uncertainty, and thermal contraction. Even more detrimental is the possibility that the measured spectrum is missing a resonance, rendering the entire analysis incorrect. While we are confident in our analysis for the particularly large and well-oriented sample S1, large samples of \urusi are known to be of slightly lower quality \cite{baumbachPhilMag}. Smaller, higher-quality crystals of \urusi do not lend themselves well to RUS studies, being hard to align and polish to high precision. Smaller samples also produce weaker RUS signals, making it easier to miss a resonance. We have therefore developed a new method for extracting symmetry information directly from the resonance spectrum, without needing to first extract the elastic moduli themselves, even if the spectrum is incomplete. This method takes advantage of the power of machine learning algorithms to recognize patterns in complex data sets.

\subsection*{Machine Learning for Resonant Ultrasound Spectroscopy}

Artificial neural networks (ANNs) are popular machine learning tools due to their ability to classify objects in highly non-linear ways. In particular, ANNs can approximate smooth functions arbitrarily well \cite{Cjaji:2001}. Here we train an ANN to learn a function that maps the jumps in ultrasonic resonances at a phase transition to one of two classes, corresponding to either a one-component or two-component OP. one-component OPs induce jumps only in compressional elastic moduli, whereas two-component OPs also induce jumps in two of the shear moduli. Phase transitions with two-component OPs should therefore show jumps in more ultrasonic resonances at a phase transition than phase transitions with one-component OPs. Our intent is that this difference in the distribution of jumps can be learned by an ANN to discriminate between one-component and two-component OPs.

 An ANN must be trained with simulated data that encompasses a broad range of possible experimental scenarios. In our case we simulate RUS spectra given assumptions about the sample and the OP dimensionality. Starting with a set of parameters randomly generated within bounds that we specify---these include the sample geometry, density, and the six elastic moduli---we solve the elastic wave equation to produce the first $N$ resonance frequencies that would be measured in an RUS experiment. Then using a second set of assumptions---whether the OP has one-component or two, whether our simulated experiment has $k$ missing resonances, and the relative sizes of the elastic constant jumps produced at \THO ---we calculate the jumps at \THO for the first $n$ resonances(see Fig. 3). By varying the input assumptions we produce a large number of training data sets that are intended to encompass the (unknown) experimental parameters. 

While the sample geometry, density, and moduli are well determined for S1 and only varied by a few tens of percent, the dimensionality of the OP, the number of missing resonances, and the sizes of the jumps in each symmetry channel are taken to be completely unknown. We vary these latter parameters across a broad range of physically possible values (see Fig. 3 and SI for further details). To prepare the simulated data for interpretation by our ANN we take the first $n$ jumps, sort the jumps by size, normalize the jumps to lie between zero and one, and label the data sets by the dimensionality of the OP that was used to create them---either one-component or two.

This normalized and sorted list of numbers $\{\Delta f_i / f_i\}$ is used as input to an ANN. Our ANN architecture is a fully connected, feed forward neural network with a single hidden layer containing 20 neurons (see Fig. 3).
Each neuron $j$ processes the inputs $\{\Delta f_i / f_i\}$ according to the weight matrix $w_{ji}$ and the bias vector $b_j$ specific to that neuron as $ \sigma(w_{ji}x_i + b_j)$ where the rectified linear activation function is given by $\sigma(y)\equiv\rm{max}(y,0)$. The sum of the neural outputs is normalized via a softmax layer. 

We train the ANN using 10000 sets of simulated RUS data for the case of a one-component OP, with varied elastic constants, sample geometries, jump magnitudes, and missing resonances, and another 10000 sets for the case of a two-component OP. We use cross-entropy as the cost function for stochastic gradient descent. We train 10 different neural networks in this way to an accuracy of $\sim 90 \%$, and then fix each individual network's weights and biases. Once the networks are trained we ask each ANN for its judgment on the OP dimensionality associated with an experimentally determined set of 29 jumps, and average the responses from each neural network. The sizes of the jumps depend on how \THO is assigned---assigning \THO artificially far from the actual phase transition will produce large jumps in all resonances. We therefore repeat our ANN determination using a range of \THO around the phase transition, and plot the outcome as a function of \THO. 
 
Fig. 4a shows the results of our ANN analysis for sample S1---the same sample discussed above using the traditional analysis. To visually compare the training and experimental data in a transparent fashion, we plot the list of sorted and normalized jumps against their indices in the list. The average of the one-component training data is shown in red; the average of the two-component training data is shown in blue; the experimental jumps are shown in grey. It is clear that the experimental data resembles the one-component training data much more closely. This resemblance is quantified in the inset, showing the ANN confidence that the experimental data belongs to the one-component class for varying assignments of \THO. We find that the confidence of a one-component OP is maximized in the region of assigned \THO that corresponds to the experimental value of \THO.

\begin{figure*}
	\centering
	\includegraphics[width=.99\linewidth]{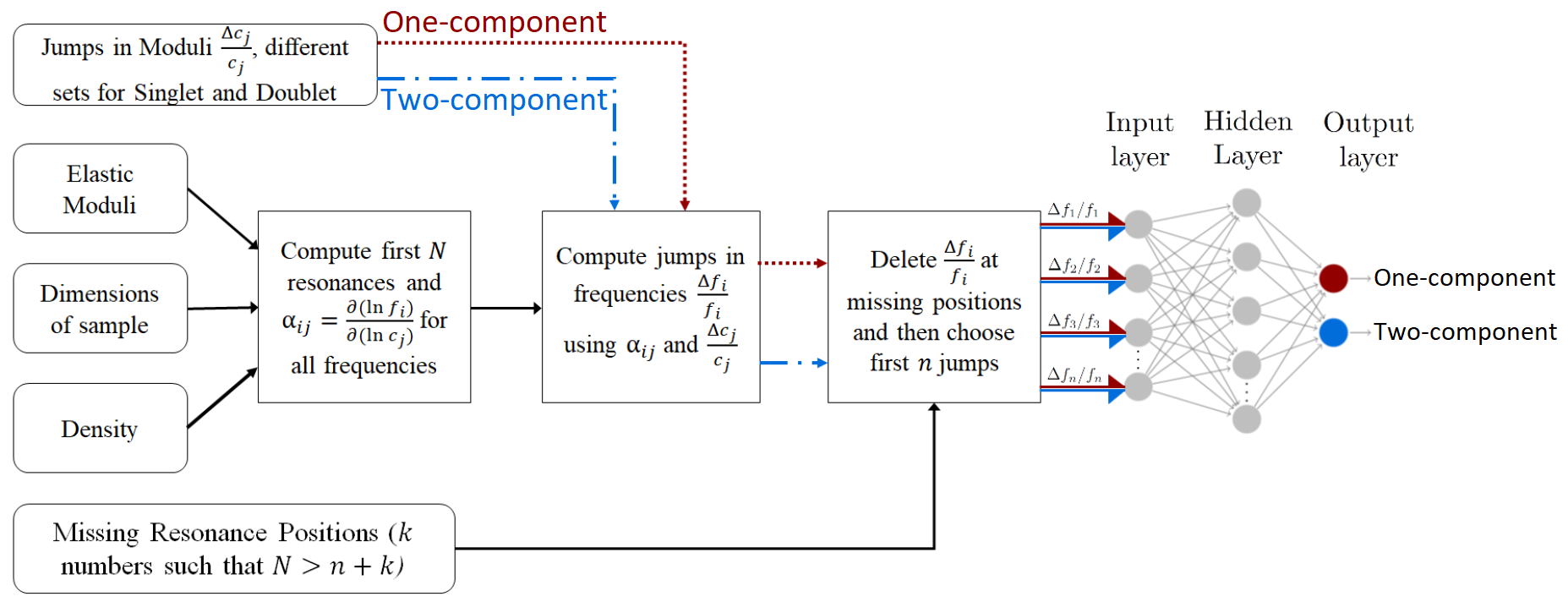}
	\caption{\textbf{Schematic of the algorithm used to generate the training data.} Values for elastic moduli and dimensions are chosen randomly from a range that bounds our experimental uncertainties. One-component OPs give jumps only in $A_{1g}$ moduli, whereas two-component OPs also give jumps in $B_{1g}$ and $B_{2g}$ moduli. Separate output files are generated corresponding to one-component and two-component OPs, each containing $n$ jumps, where $n$ is the number of frequencies whose temperature evolution could be experimentally measured. We use scaled RUS frequency shifts $\Delta f_j / f_j$ as input to the ANN. The neurons in the hidden layer have weights $w_{ij}$ and biases $b_i$. Each output neuron corresponds to one of the two OP dimensionalities under consideration, i.e. one-component and two-component. The output value of each neuron is the network's judgment on the likelihood of that OP dimensionality.}
	\label{fig:nn_diagram}
\end{figure*}

\begin{figure*}
	\label{results}
	\includegraphics[width=0.99\linewidth]{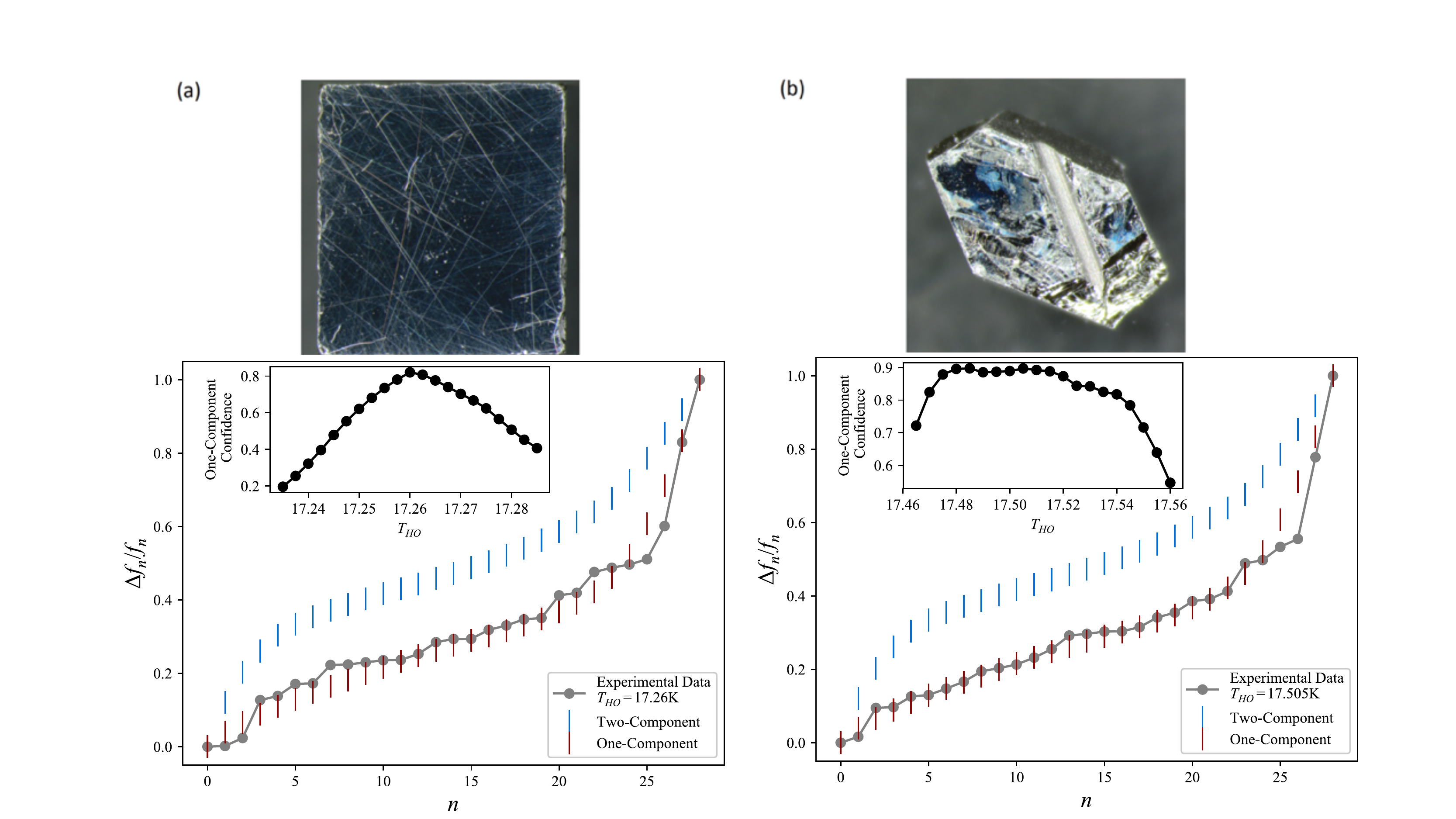}
	\caption{\textbf{Results of the ANN analysis for two samples of \urusi.} Upper blue curves show the averaged, sorted, simulated frequency shift (``jump'') data plotted against its index in the sorted list for a two-component OP for     (a) sample S1 and (b) sample S2. 
		The data are normalized to range from 0 to 1. Lower, red curves shows the same for a one-component OP. Grey dots show experimental data for critical temperature assignment (a) \THO $= 17.26$ K and (b) \THO $= 17.505$ K, which visually aligns more closely with the average one-component simulated data than the two-component simulations. Insets: percent confidence of the one-component output neuron for various assignments of \THO averaged over 10 trained networks. 
		A maximum confidence of (a) $83.2\%$ occurs for \THO $= 17.26$ K and 
		(b) $89.7\%$ for \THO $= 17.505$ K. Sample S2 has a higher value of \THO due to its lower impurity concentration, as verified independently by the resistivity. Photo credits: Sayak Ghosh, Cornell University.}
	\label{fig:results}
\end{figure*}

Thus far we have shown that both methods---the traditional method of extracting the elastic moduli using the elastic wave equation, and our new method of examining the resonance spectrum directly using a trained ANN---agree that the HO parameter of \urusi is one-component. We can now use the neural network to analyze a smaller, irregular-shaped but higher quality (higher \THO \cite{baumbachPhilMag}) sample that cannot be analyzed using the traditional method due to its complicated geometry. Figure 4b shows the result of ANN analysis performed on a resonance spectrum of sample S2. The sorted and normalized spectrum looks very similar to that of S1, and the averaged ANN outcome gives 90\% confidence that the OP is one-component. Despite the fact that S2 has a geometry such that the elastic moduli cannot be extracted, its resonance spectrum still contains information about the OP dimensionality and our ANN identifies this successfully.

\section*{Discussion}

Our two analyses of ultrasonic resonances across \THO in \urusi strongly support one-component OPs, such as electric-hexadecapolar order \cite{haule:np2009a}, the chiral density wave observed by Raman spectroscopy \cite{Buhot:2014,Kung:2015,Kung:2016}, and are consistent with the lack of $C_4$ symmetry breaking observed in recent X-ray scattering experiments \cite{Choi:2018}. Our analysis rules against two-component OPs, such as rank-5 superspin \cite{rau:pr2012a,kambe:pr2018a} and spin-nematic order \cite{fujimoto:prl2011a}. The power of our result lies in its independence from the microscopic origin of the OP: group-theoretical arguments alone are sufficient to rule out large numbers of possible OPs. It could be argued that the coupling constants governing the jumps in the shear moduli are sufficiently small such that the jumps are below our experimental resolution. Previous experiments, however, have shown these coupling constants to be of the same order of magnitude in other materials with multi-component OPs \cite{LuthiPRB,IkedaStSol,GarlandJChemPhys}. It has also been demonstrated that the size of the jump in heat capacity at \THO is largely insensitive to RRR\cite{Matsuda:2011, baumbachPhilMag,Gallagher:2016}. It is therefore hard to imagine that higher RRR samples would yield jumps in the shear moduli.

The use of ANNs to analyze RUS data represents an exciting opportunity to re-examine ultrasound experiments that were previously unable to identify order parameter symmetry. For example, irregular sample geometry prevented identification of the order parameter symmetry in the high-T$_{\mathrm{c}}$ superconductor YBa$_2$Cu$_3$O$_{6.98}$ \cite{Shekhter:2013}. Re-analysis of this spectrum using our ANN could reveal whether the OP of the pseudogap is associated with $E_u$-symmetry orbital loop currents. The proposed two-component $p_x + i p_y$ superconducting state of Sr$_2$RuO$_4$ and other potential spin-triplet superconductors could also be identified in this fashion, where traditional pulse-echo ultrasound measurements have been confounded by systematic uncertainty \cite{Okuda:2002}. 

Beyond RUS, there are many other data analysis problems in experimental physics that stand to be improved using an approach similar to the one presented here \cite{Greitemann:2019}. In particular, any technique where simulation of a data set is straightforward but where fitting is difficult should be amenable to a framework of the type used here. The most immediately obvious technique where our algorithm could be applied is NMR spectroscopy. NMR produces spectra in a similar frequency range to RUS, but which originate in the spin-resonances of nuclear magnetic moments. Modern broad-band NMR can produce complex temperature-dependent spectra, containing resonances from multiple elements situated at different sites within the unit cell. Given a particular magnetic order it is relatively straightforward to calculate the NMR spectrum---i.e. to produce training data. The inverse problem, however, is more challenging: recovering a temperature-dependent magnetic structure from an NMR data set. In a way similar to RUS, missing resonances, and resonances mistakenly attributed to different elements, can render an analysis entirely invalid. It should be relatively straightforward to adapt our framework for generating training data and our ANN to extract temperature (or magnetic field) dependent magnetic structures from NMR spectra.

\section*{Acknowledgements}
We thank P. Coleman, P. Chandra, and R. Flint for helpful discussions.

\section*{Funding}
Work at Los Alamos National Laboratory was performed under the auspices of the US Department of Energy, Office of Basic Energy Sciences, Division of Materials Sciences and Engineering. M.M., B.J.R., and S.G. acknowledges support by the Cornell Center for Materials Research with funding from the NSF MRSEC program (DMR-1719875). M.M. acknowledges support by the National Science Foundation (Platform for the Accelerated Realization, Analysis, and Discovery of Interface Materials (PARADIM)) under Cooperative Agreement No. DMR-1539918. E.-A.K. acknowledge support from DOE DE-SC0018946. B.J.R. and S.G. Acknowledge funding from the National Science Foundation under Grant No. DMR-1752784.

\section*{Author Contributions}
S.G. and B.J.R designed the experiment; R.B., E.D.B. grew sample S2; J.A.M. grew sample S1; S.G. acquired and analyzed the data; M.M. and E.A.K. designed the ANN; and K.A.M., A.S., M.M., S.G., and B.J.R. wrote the manuscript with input from all co-authors. 

\section*{Data Availability}
All data needed to evaluate the conclusions in the paper are present in the paper and/or the Supplementary Materials. Additional data available from authors upon request.

\section*{Materials and Methods}

Sample S1 was grown by the Czochralski method. A single crystal oriented along the crystallographic axes was polished to dimensions 3.0mm $\times$ 2.8mm $\times$ 2.6mm, with 2.6mm along the tetragonal long axis. Sample S2 was grown was grown by the Czochralski method and then processed by solid-state electrorefinement. Typical RRR values for ab-plane flakes of \urusi taken from the larger piece range from 100-500. The RRR values measured on larger pieces (Fig. 4) are between 10-20. For a comparison of different growth methods for \urusi see Gallagher et al. \cite{Gallagher:2016}.

Resonant ultrasound experiments were performed in a custom-built setup consisting of two compressional-mode lithium niobate transducers, which were vibrationally isolated from rest of the apparatus. The top transducer was mounted on a freely pivoting arm, ensuring weak coupling and linear response. The response voltage generated on the pickup transducer---maximum whenever the drive frequency coincides with a sample resonance---was measured with lock-in technique. The response signal is preamplified using a custom-made charge amplifier to compensate for signal degradation in coaxial cables \cite{miglioriRSI}. Oxford Instruments $\mathrm{He^4}$ cryostat was used for providing temperature control.

\section*{Supplementary Information}
Phase-locked Loop\\
Training data for ANN\\
Symmetry and Coupling\\
Lack of c$_{33}$ jump\\
Resolving the origin of jumps\\
Compositions of Resonances\\
Resistance Measurement\\
Possible Effects from Parasitic AFM\\
Table S1. Calculated discontinuities(``jumps") in elastic moduli for one- and two-component order parameters in a tetragonal system.\\
Fig. S1. Resonant ultrasound using phase-locked loop.\\
Fig. S2. Three representative resonance frequencies of URu$_2$Si$_2$ and their attenuation through T$_{HO}$.\\
Fig. S3. Elastic moduli of URu$_2$Si$_2$ with the contribution above T$_{HO}$ subtracted.\\
Fig. S4. Fitting temperature evolution of resonances.\\
Fig. S5. Resistance of sample S2 measured from 300K down to 2K.

\section*{Competing Interests} The authors declare that they have no competing interests.

\nocite{Altmann:1994}
\nocite{RehwaldAIP1973}
\nocite{MiyakePRL1986}
\nocite{MatsudaPRL2001}
\nocite{Yanagisawa2014}
\bibliography{library}

\bibliographystyle{Science}
\newpage
\input{Combined_SI_v2.tex}
\end{document}

%% file: Combined_SI_v2.tex
 
\section*{Supplementary Information} 

    \subsection*{Phase-locked Loop}

\begin{figure}
		\centering
		\includegraphics[width=\linewidth]{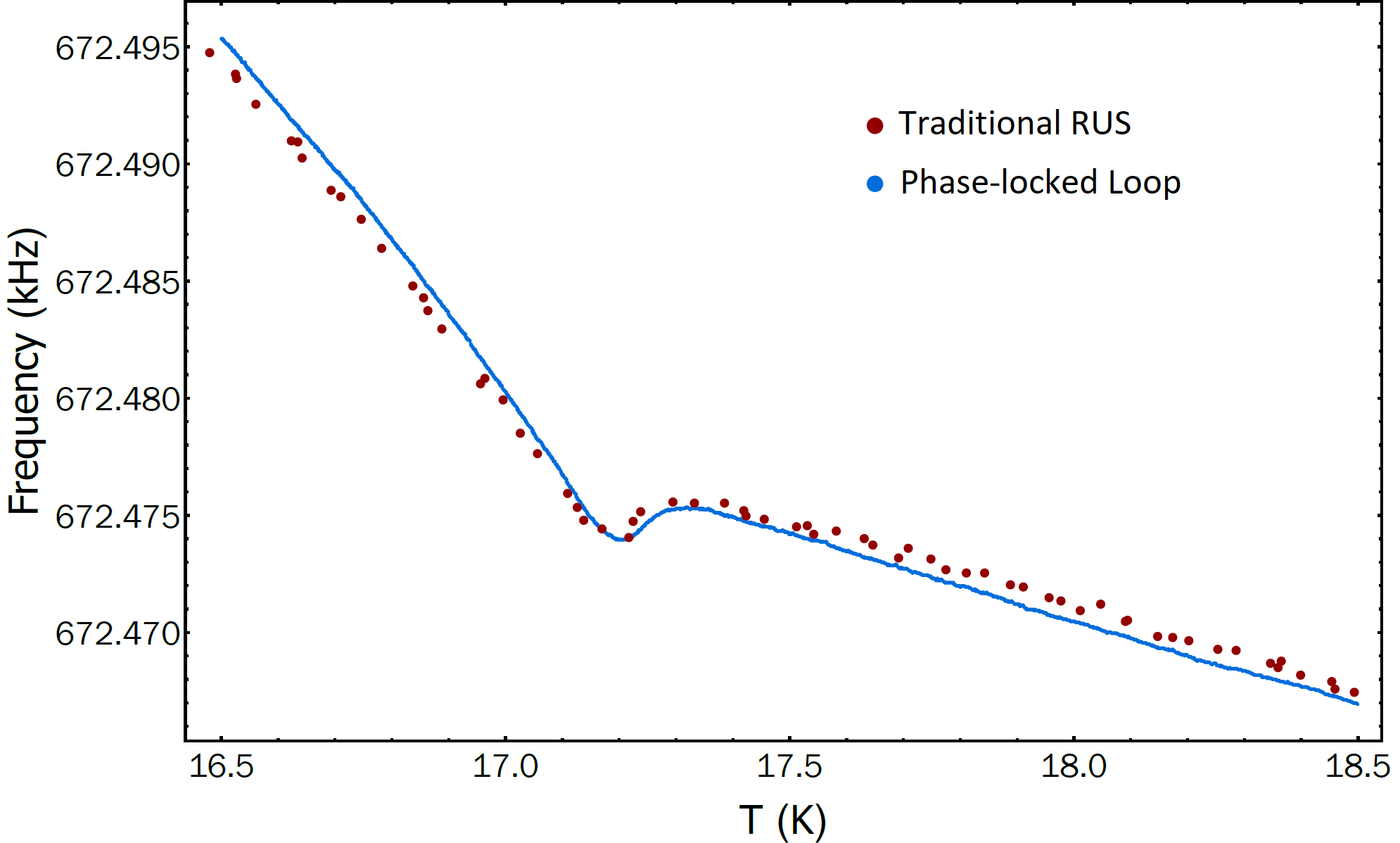}
		\caption{Resonant ultrasound using phase-locked loop. We compare data between fitting Lorentzian(traditional RUS) and phase-locked loop(PLL) for the same resonance frequency as sample is cooled through the hidden order transition. Data taken with PLL has lower fluctuations and a much higher density of frequency points.}
		\label{fig:comparison}
	\end{figure}

    The traditional method to track resonance frequencies with temperature in RUS is to fit a Lorentzian to find the resonance, and this requires measuring the spectra in a small frequency range around the resonance at every temperature. For our RUS measurements, we use a phase-locked loop (PLL) to track resonance frequencies as a function of temperature, which eliminates the need to scan over a frequency range at every temperature. PLL comprises a lock-in amplifier coupled to a PID controller. Lock-in measures amplitude and phase of the response at the drive frequency, and the PID controller adjusts drive frequency so as to maintain the phase at a setpoint. We typically choose the phase setpoint to coincide with the peak of the resonance since the phase changes most rapidly there. We find that compared to traditional RUS, PLL tracks resonances with higher sensitivity and the standard error is lower by a factor of 30 (see Figure \ref{fig:comparison}). As there is no need to do a frequency scan and perform a Lorentzian fit at every temperature, this is also a faster measurement yielding $\sim$1000 times more points per unit temperature. To the best of our knowledge, a phase-locked loop technique has never been employed before in resonant ultrasound spectroscopy studies.
    
    \subsection*{Training data for ANN}
    We generate simulated RUS spectra for training the artificial neural network (ANN). We first solve the 3-D elastic wave equation using input parameters---density of sample, elastic constants and dimensions---which are chosen from a range that bounds our experimental uncertainties, to obtain the first $N$ lowest frequency resonances and their compositions ($\alpha_{ij}=\partial(\ln{f_i})/\partial(\ln{c_j})$) in terms of the six independent elastic constants. We expect different sets of jumps in moduli ($\Delta c_j/c_j$) for one- and two-component OPs (see `Symmetry and Coupling' section for details), which translate into different sets of jumps in resonance frequencies($\Delta f_i/f_i$) for the two OP types. Experimentally, some resonances are too noisy to be measured through the transition, and to account for that, we delete a random number ($k$) of jumps from the the list of $N$ consecutive $\Delta f_i / f_i$. We then choose the first $n$ jumps to constitute a training data set, where $n$ is the number of jumps measured in experiment.

    \subsection*{Symmetry and Coupling}
	In a tetragonal crystal, elastic strain breaks into five irreducible representations, two compressive strains transforming as the $A_{1g}$ representation and three shear strains
	transforming as the $B_{1g}$, $B_{2g}$ and $E_{g}$ representations [3]. The elastic free energy density is given by
		\begin{equation}
		\begin{aligned}[b]
		\mathcal{F}_{el} &=\frac{1}{2}\Big(c_{11}(\epsilon_{xx}^2+\epsilon_{yy}^2)+2c_{12}\epsilon_{xx}\epsilon_{yy}+c_{33}\epsilon_{zz}^2+2c_{13}(\epsilon_{xx}+\epsilon_{yy})\epsilon_{zz}+4c_{44}(\epsilon_{xz}^2+\epsilon_{yz}^2)+4c_{66}\epsilon_{xy}^2\Big) \\
		&=\frac{1}{2}\Big(\frac{c_{11}+c_{12}}{2}(\epsilon_{xx}+\epsilon_{yy})^2+c_{33}\epsilon_{zz}^2+2c_{13}(\epsilon_{xx}+\epsilon_{yy})\epsilon_{zz}+\frac{c_{11}-c_{12}}{2}(\epsilon_{xx}-\epsilon_{yy})^2+\\&4c_{44}(\epsilon_{xz}^2+\epsilon_{yz}^2)+4c_{66}\epsilon_{xy}^2\Big) \\
		&=\frac{1}{2}\Big(c_{A_{1g,1}}\epsilon_{A_{1g,1}}^2+c_{A_{1g,2}}\epsilon_{A_{1g,2}}^2+2c_{A_{1g,3}}\epsilon_{A_{1g,1}}\epsilon_{A_{1g,2}}+c_{B_{1g}}\epsilon_{B_{1g}}^2+c_{E_g}|\epsilon_{E_g}|^2+c_{B_{2g}}\epsilon_{B_{2g}}^2\Big)
		\end{aligned}
		\label{eqn:Felastic}
		\end{equation}
	where the strains are written as the irreducible representations of $D_{4h}$, $(\epsilon_{xx}+\epsilon_{yy}) \rightarrow \epsilon_{A_{1g,1}}$, $\epsilon_{zz} \rightarrow \epsilon_{A_{1g,2}}$, $(\epsilon_{xx}-\epsilon_{yy}) \rightarrow \epsilon_{B_{1g}}$, $2\epsilon_{xy} \rightarrow \epsilon_{B_{2g}}$ and $(2\epsilon_{xz},2\epsilon_{yz}) \rightarrow \epsilon_{E_g}$.
	
	We consider integer representations of $D_{4h}$ in the following calculations. Half-integer representations have been addressed at the end of this section. 
	
	Following Landau theory, near a phase transition, an order parameter(OP) $\eta$ is introduced whose contribution to the free energy density is 
	\begin{equation}
	 \mathcal{F}_{0}^{s}(\eta) =a\eta^2+\frac{b}{2}\eta^4
	 \end{equation}
	For a two-component OP $(\eta_x,\eta_y)$, this becomes 	
	\begin{equation}
	\mathcal{F}_{0}^{d}(\eta_x,\eta_y) =a(\eta_x^2+\eta_y^2)+\frac{b_0}{4}(\eta_x^2+\eta_y^2)^2+\frac{b_1}{4}(\eta_x^2-\eta_y^2)^2+b_2\eta_x^2\eta_y^2
	\end{equation}
	with $a=a_0(T-T_{HO})$, and $a,b,b_{0,1,2}>0$. 
	With strain-OP coupling($\mathcal{F}_c$), the total free energy density becomes
	\begin{equation}
	\mathcal{F}=\mathcal{F}_0(\eta_i) + \mathcal{F}_{el}(\epsilon_{\mu}) + \mathcal{F}_c(\eta_i,\epsilon_\mu)
	\end{equation} 
	
	The coupling between OP and various symmetry strains leads to the corresponding elastic constants getting modified in the ordered state. Since the free energy density must transform like the identity $A_{1g}$ representation, only those couplings are allowed in $\mathcal{F}_c$ which transform as $A_{1g}$ under symmetry operations of the tetragonal($D_{4h}$) group. 
	
	The square of any one-component order parameter has $A_{1g}$ symmetry, and hence an one-component OP can couple to only $A_{1g}$ strains. A two-component OP in a tetragonal crystal belongs to $E_{g/u}$ representation and since the square of an $E_{g/u}$ object has $A_{1g}$, $B_{1g}$ and $B_{2g}$ symmetry objects, we expect a two-component OP to couple to $\epsilon_{B_{1g}}$  and $\epsilon_{B_{2g}}$ in addition to the two $A_{1g}$ strains. None of the OPs can couple to $\epsilon_{E_{g}}$. 
	
	Written explicitly, the coupling free energy reads
	\begin{equation}
	\mathcal{F}_{c}^s =(g_1\epsilon_{A_{1g,1}}+g_2\epsilon_{A_{1g,2}})\eta^2
	\end{equation}
	for one-component OPs and 
	\begin{equation}
	\mathcal{F}_{c}^d =(g_1\epsilon_{A_{1g,1}}+g_2\epsilon_{A_{1g,2}})(\eta_x^2+\eta_y^2)+g_4\epsilon_{B_{1g}}(\eta_x^2-\eta_y^2)+\frac{g_5}{2}\epsilon_{B_{2g}}\eta_x\eta_y
	\end{equation}
	for two-component OPs in a tetragonal system. Clearly, for non-zero discontinuity(``jump") in a particular moduli $c_{\mu}$, there needs to be an allowed coupling term between the associated symmetry strain $\epsilon_\mu$ and the order parameter.
	
	 \begin{table*}
	 	\renewcommand{\arraystretch}{2}
	 	\begin{tabular}{||c|c|| c| c|c|c|c||}
	 		\hline
	 		Dimensionality & Symmetry & $\Delta{c_{A_{1g},1}}$ & $\Delta{c_{A_{1g},2}}$ & $\Delta{c_{A_{1g},3}}$ & $\Delta{c_{B_{1g}}}$ & $\Delta{c_{B_{2g}}}$ \\ 
	 		\hline \hline
	 		One-component & A$_1$, A$_2$, B$_1$, B$_2$ & \large{$\frac{-g^2_1}{b}$} &\large{$\frac{-g^2_2}{b}$} & \large{$\frac{-g_1g_2}{b}$}& 0 & 0 \\
	 		\hline 
	 		\multirow{2}{3.5em}{Two-component} &E$(\pm1,0)$,E$(0,\pm1)$ & \large{$\frac{-2g^2_1}{(b_0+b_1)}$} & \large{$\frac{-2g^2_2}{(b_0+b_1)}$} & \large{$\frac{-2g_1g_2}{(b_0+b_1)}$} & \large{$\frac{-2g^2_4}{(b_0+b_1)}$} & \large{$\frac{-2g^2_5}{(b_2-b_1)}$} \\
	 		\cline{2-7}
	 		& E$(\pm 1, \pm 1)$ &  \large{$\frac{-2g^2_1}{(b_0+b_2)}$} & \large{$\frac{-2g^2_2}{(b_0+b_2)}$}& \large{$\frac{-2g_1g_2}{(b_0+b_2)}$} & \large{$\frac{-2g^2_4}{(b_1-b_2)}$} & \large{$\frac{-2g^2_5}{(b_0+b_2)}$} \\
	 		\hline	
	 	\end{tabular}
	 	\caption{\label{tab:table2} Calculated discontinuities(``jumps") in elastic moduli for one- and two-component order parameters in a tetragonal system.}
	 	\label{tab:jumps}
	 \end{table*}

	As a concrete example, we consider a two-component OP $(\eta_x,\eta_y)$ of $E$-symmetry and calculate jumps in the various moduli. Minimizing $\mathcal{F}_{0}^{d}$ gives two distinct equilibrium OPs depending on the relative magnitudes of $b_1$ and $b_2$. 
	\begin{equation}
	(\eta_{x}^0,\eta_{y}^0) = \begin{cases}
	\sqrt{-\frac{2a}{b_0+b_1}}\{(\pm 1,0) \:, (0,\pm 1)\}  & ,b_1<b_2\\
	\sqrt{-\frac{a}{b_0+b_2}}(\pm 1,\pm 1) & ,b_1>b_2
	\end{cases}
	\end{equation}
	
	For $b_1<b_2$, the 4 possible states(all of which are physically equivalent) are denoted as $E(\pm1,0)$ for $\eta_y^0=0$ and $E(0,\pm1)$ for $\eta_x^0=0$. For $b_1>b_2$, the possible equivalent states are denoted as $E(\pm1,\pm1)$. To calculate strain-dependent corrections to the equilibrium OP $(\eta_{x}^0,\eta_{y}^0)$, we perturb $(\eta_{x}^0,\eta_{y}^0)\rightarrow(\eta_{x}^0+\tilde{\eta_x},\eta_{y}^0+\tilde{\eta_y})$ and expand the $(\eta_{x},\eta_{y})$-dependent terms in $\mathcal{F}$ up to quadratic order in $(\tilde{\eta_{x}},\tilde{\eta_{y}})$. Minimizing with respect to $\tilde{\eta_x}$ and $\tilde{\eta_y}$ then reduces to solving two linear equations, which give us the modified OP. Finally, replacing this OP in $\mathcal{F}$ and taking appropriate strain derivatives allows to straightforwardly calculate the modified elastic constants in the ordered state.	
	
	The calculated jumps in the various moduli for all the integer representation OPs of $D_{4h}$ are tabulated in Supplementary Table \ref{tab:jumps}. Clearly, a two-component OP induces jumps in both of the in-plane shear($B_{1g}$ and $B_{2g}$) moduli in addition to the compressional($A_{1g}$) moduli, while an one-component OP only modifies the $A_{1g}$ moduli, as expected. This leads to different sets of $\Delta c_j /c_j$ for one-component and two-component OPs, which we use to generate training data for the neural network (detailed in Figure 3 of the main text and the `Training data for ANN' section). We note that the non-zero shear jumps concluded from our calculations refute the claim made in [20] that shear moduli are continuous through the hidden order transition.
	
	$D_{4h}$ also contains half-integer representations, such as $E_{3/2,g}$ which is the representation of the slave boson responsible for the ``hastatic'' order parameter proposed by Chandra \textit{et al.} [35] ($E_{3/2,g}$ is also known as $\Gamma_7^+$).  $E_{3/2,g}$ forms bilinears of the $A_{2g}$ and $E_g$ representations (see Altmann and Herzig [55] table 33.8): this predicts no jump in either $c_{66}$ or $(c_{11}-c_{12})/2$, but does predict a jump in $c_{44}$, which we also do not observe. If, however, the hastatic order only forms bilinears at finite $\textbf{q}$, then it will predict no jumps in any of the elastic moduli.

	\subsection*{Lack of $\mathbf{c_{33}}$ jump}
	
	In our experiment, we do not see a discontinuity in $c_{33}$ elastic constant, although it is expected for an one-component OP. From Supplementary Table \ref{tab:jumps}, the three $A_{1g}$ jumps can be related as 
	\begin{equation}
	\Delta{c_{A_{1g},1}} \times \Delta{c_{A_{1g},2}} =(\Delta{c_{A_{1g},3}})^2
	\label{eqn:a1gjumps}
	\end{equation} 
	Experimentally we measure relative jumps($\frac{\Delta{c}}{c}$), which can be related by rewriting \autoref{eqn:a1gjumps} as
	\begin{equation}
	\begin{aligned}[b]
	&\frac{\Delta{c_{A_{1g},2}}}{c_{A_{1g},2}}\cdot c_{A_{1g},2} =\frac{\Big(\frac{\Delta{c_{A_{1g},3}}}{c_{A_{1g},3}}\cdot c_{A_{1g},3}\Big)^2}{\Big(\frac{\Delta{c_{A_{1g},1}}}{c_{A_{1g},1}}\cdot c_{A_{1g},1}\Big)}\\
	&\implies \bigg(\frac{\Delta{c_{A_{1g},2}}}{c_{A_{1g},2}}\bigg)=\frac{1}{c_{A_{1g},2}} \cdot \frac{\Big(\frac{\Delta{c_{A_{1g},3}}}{c_{A_{1g},3}}\cdot c_{A_{1g},3}\Big)^2}{\Big(\frac{\Delta{c_{A_{1g},1}}}{c_{A_{1g},1}}\cdot c_{A_{1g},1}\Big)}
	\end{aligned}
	\end{equation} 
	Using the known jumps in $(c_{11}+c_{12})/2$ and $c_{23}$ (see Figure 2 of the main text and \autoref{eqn:Felastic}), we estimate $\frac{\Delta{c_{33}}}{c_{33}}\approx4\times10^{-7}$, which is an order of magnitude below our experimental resolution. This explains the lack of $c_{33}$ jump in our experimental data.
	
	\subsection*{Resolving the Origin of Jumps}
	
	The elastic moduli discontinuities predicted by $g\eta^2\epsilon$ coupling in Landau theory only show up in experiments under the assumption that the applied strain varies at a timescale much longer than the relaxation times of the order parameter [56]. Our experiments are in the low MHz frequency range, corresponding to timescales of order $10^{-7}$s. The HO transition is a mean-field like second order transition, and the OP relaxation timescales for similar (superconducting) transitions in other uranium-based heavy fermion compounds are of the order $10^{-10}-10^{-12}$s [57]. Thus our experimental frequencies are low enough to observe the moduli jumps. 
	
\begin{figure}
		\centering
		\includegraphics[width=.99\linewidth]{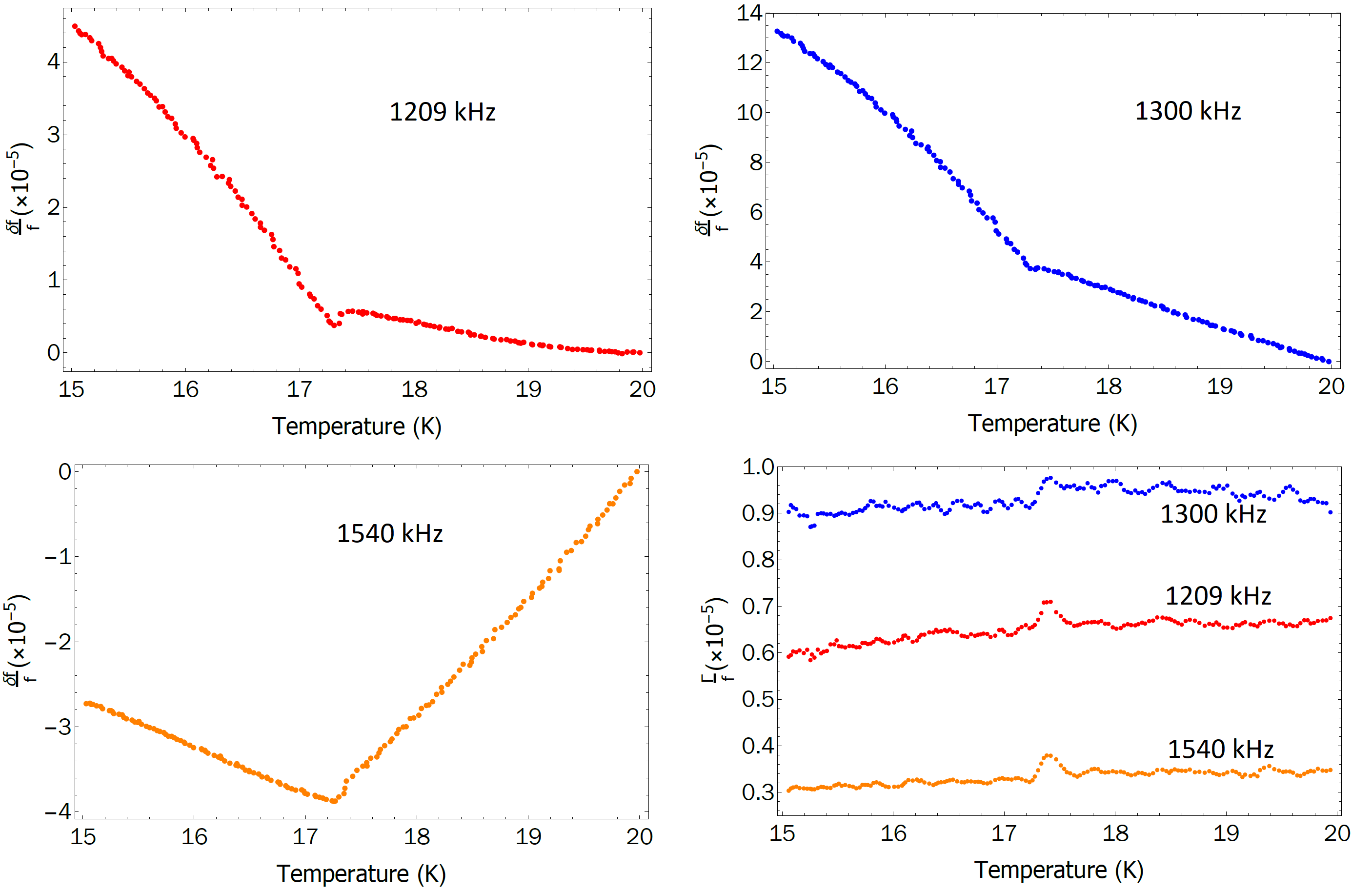}
		\caption{Three representative resonance frequencies of URu$_2$Si$_2$ and their attenuation through T$_{HO}$. A clear discontinuity can be seen in all three resonances and a small peak-like feature can be seen in the attenuation at T$_{HO}$. The lack of any large temperature-dependent feature in the attenuation rules out anomalous order parameter dynamics interfering with our results.}
		\label{fig:origin}
	\end{figure}

\begin{figure}
		\centering
		\includegraphics[width=.99\linewidth]{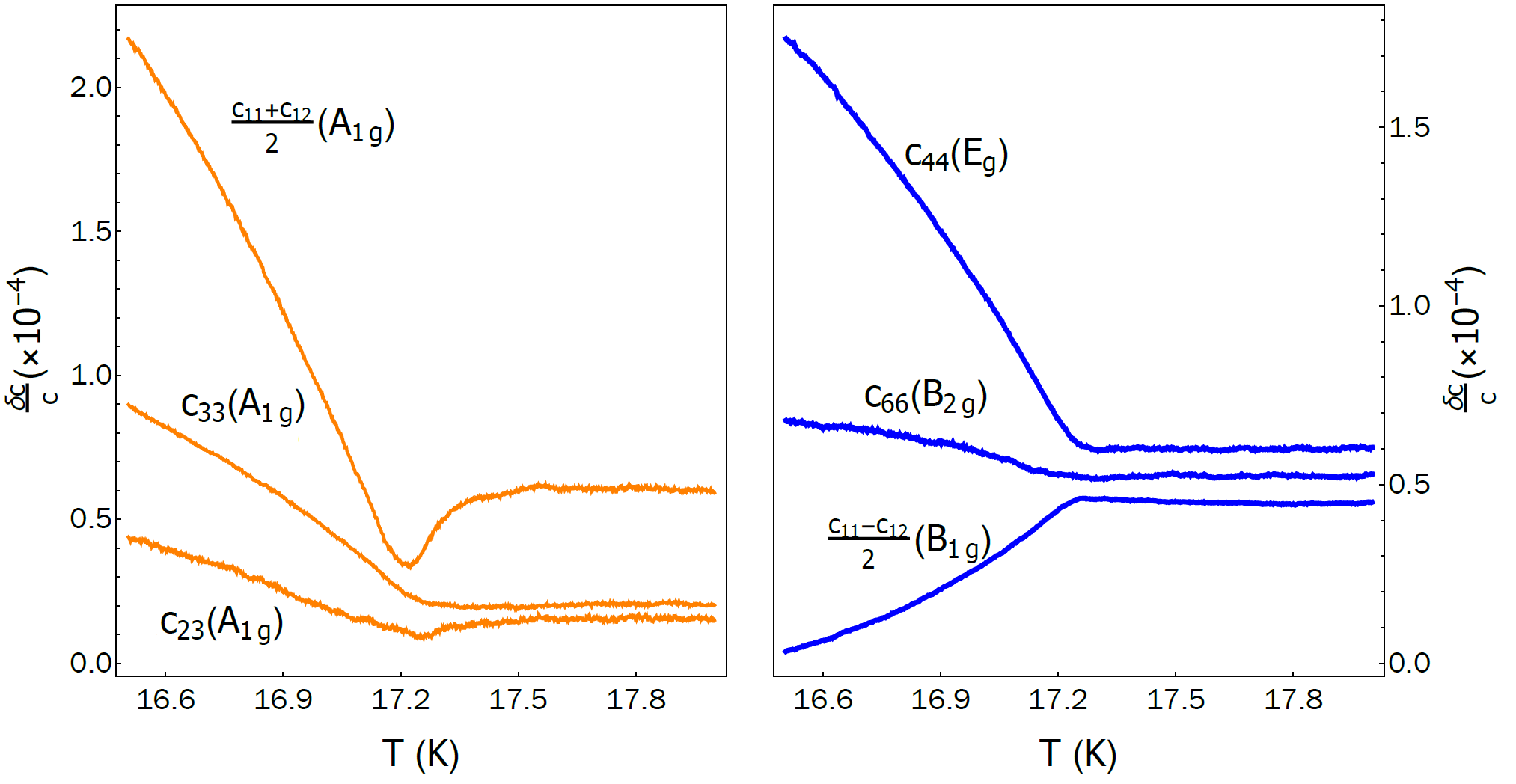}
		\caption{Elastic moduli of \urusi with the contribution above \THO subtracted. This procedure isolates the effect of the order parameter on the moduli. Plots are vertically shifted for visual clarity. It can be clearly seen that $(c_{11} + c_{12})/2$ and $c_{23}$ show a drop at \THO before starting to stiffen in the HO phase, while all other moduli only show increased stiffening/softening through \THO.}
		\label{fig:background}
	\end{figure}

	To rule out anomalous OP dynamics near the phase transition interfering with our measured jumps, we look at the ultrasonic attenuation in the resonances through \THO. In particular, anomalous OP dynamics would lead to large, temperature-dependent features in the attenuation near \THO. In Figure \ref{fig:origin}, we plot the temperature evolution of three resonances, each showing a clear discontinuity at \THO. The inverse quality factor of these frequencies (proportional to ultrasonic attenuation), plotted in panel (d), show tiny features at \THO, but overall unremarkable behavior. The tiny features are required by Kramers-Kronig relations, but the overall behavior is dominated by phonon anharmonicity and not OP dynamics.  Our plot (see Figure \ref{fig:background}) of the elastic moduli with the background contribution subtracted further highlights the singular nature of the discontinuity in the $A_{1g}$ moduli.
	
	This confirms that the moduli jumps we see originate from strain-order parameter coupling terms of the form $g\eta^2\epsilon$, and hence our measured jumps constrain the order parameter in \urusi to be one-component.

	\subsection*{Compositions of Resonances}
	
\begin{figure}
		\centering
		\includegraphics[width=\linewidth]{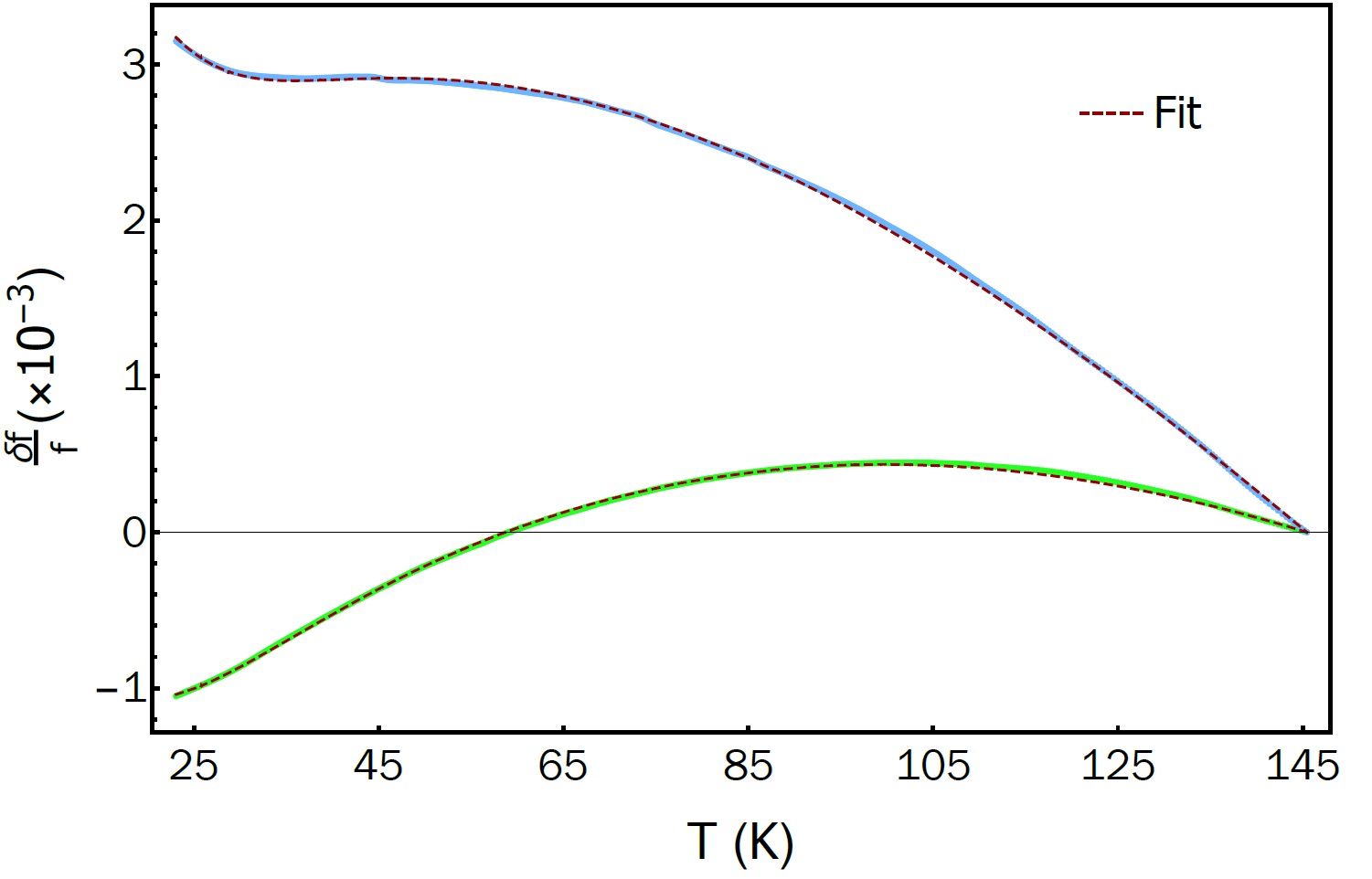}
		\caption{Fitting temperature evolution of resonances. Temperature dependence of two representative resonances(blue and green) between $\sim$25K to 145K. The fit estimates that the blue curve is composed of mostly $E_{g}$ and some $A_{1g}$ symmetry vibrations, whereas the green curve is dominated by vibrations in $B_{1g}$ channel.}
		\label{fig:fit_to_freq}
	\end{figure}

	The irregular shape of the higher quality sample S2 prevents directly inverting the frequency spectra to calculate the elastic moduli and the compositions of the frequencies in terms of the moduli ($\alpha_{ij}$s). To estimate the compositions, we use the known temperature evolution of the 6 elastic constants between 145K to 25K to fit the change in frequencies as the sample is cooled in this temperature range (Figure \ref{fig:fit_to_freq}). This helps us ensure that the sample S2 has in-plane shear ($B_{1g}$ and $B_{2g}$) modes in our experimentally accessible frequency range, and hence the data we obtain should allow us to distinguish between one- and two-component order parameters. 
	
	\subsection*{Resistance Measurement}
	
	We checked the quality of sample S2 by measuring its resistance from 300K down to 2K, as shown in Figure \ref{fig:RvT}. The Kondo crossover is seen around $\mathrm{T^*}=75$K and the resistance shows a sudden increase  at the HO transition, consistent with previous reports. We use the feature near the hidden order transition to determine \THO$=17.52$K for this sample. This value of \THO [45,51] confirms the higher quality of S2 compared to S1.

	\subsection*{Possible Effects from Parasitic AFM}
	
	Under hydrostatic pressure, \urusi undergoes a transition into a large moment antiferromagnet (LMAF) phase. Even at ambient pressure, \urusi samples are known to host some amount of this parasitic AFM phase, arising from inhomogeneous strain in the sample [58]. The AFM phase shows a large discontinuity in the modulus $c_{11}$ [59]. One could imagine that the jumps we see at T$_{HO}$ are arising from the presence of this parasitic AFM phase, in particular for sample S1, since it has a suppressed T$_{HO}=17.25$ K, compared to the standard T$_{HO}=17.5$ K. 
	
\begin{figure}
		\centering
		\includegraphics[width=\linewidth]{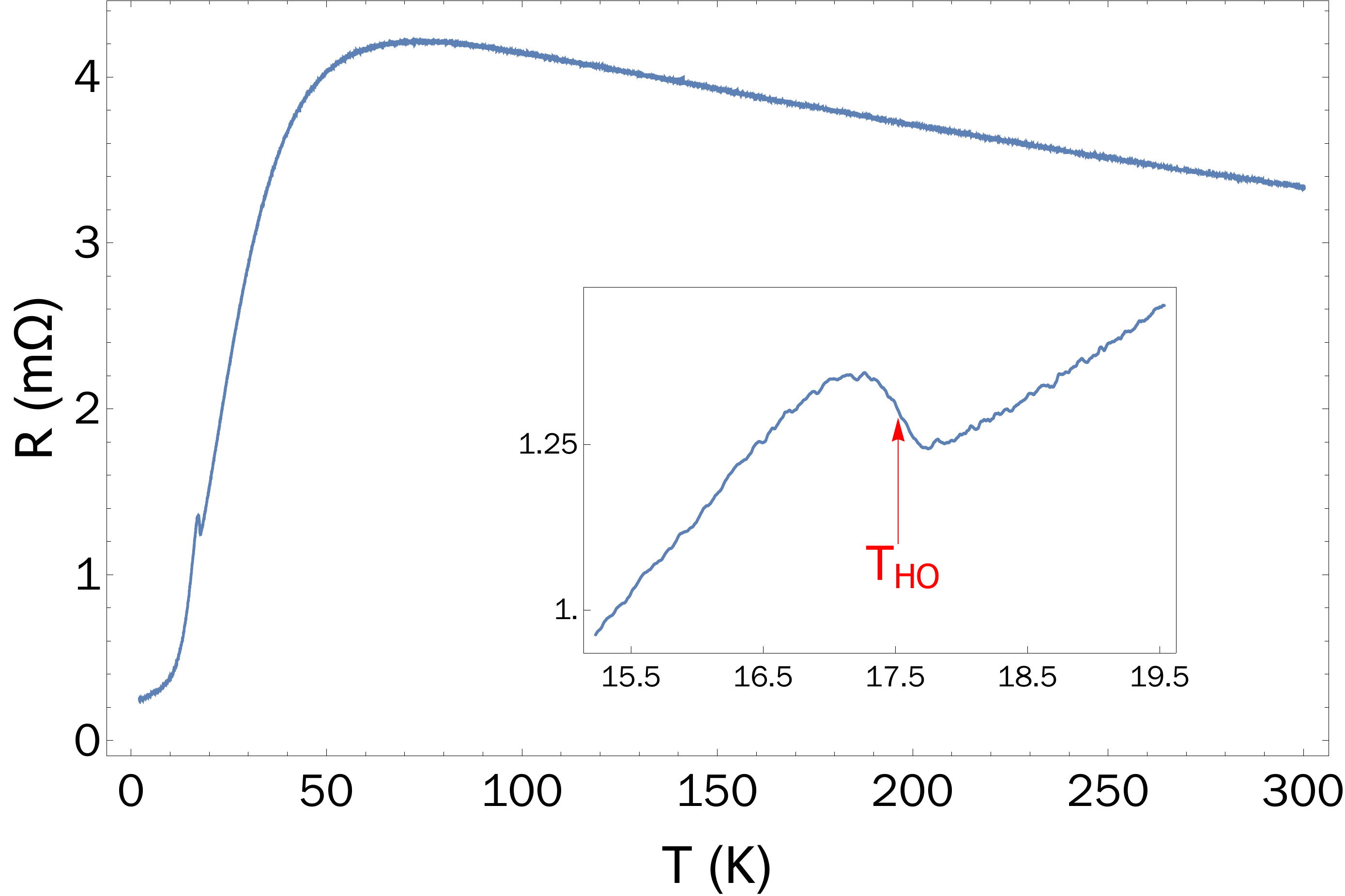}
		\caption{Resistance of sample S2 measured from 300K down to 2K. Inset shows the feature at hidden order transition, from which we determine \THO$=17.52$K for this sample.}
		\label{fig:RvT}
	\end{figure}

	To confirm that we are seeing discontinuities from the HO phase, we obtained the high-purity sample S2, which underwent solid-state electrorefinement, and has T$_{HO}=17.5$ K. It can be seen that sample S2 shows the same distribution of jumps as S1 (see Figure 4 of the main text). This rules against any parasitic effects leading to the jumps, since these samples must have very different concentrations of AFM puddles in them. Additionally, we note that the elastic anomalies due to strain-induced AFM should appear at T$_{N}$, which is a different temperature than T$_{HO}$. We see only a single transition, as evidenced by a single jump in A$_{1g}$ moduli and a change in slope in the shear moduli, all happening at the HO transition temperature (as defined by resistance measurements shown in Figure \ref{fig:RvT}). If the elastic anomalies were to arise from strain-induced AFM, we would expect these signatures of the phase transition to show up at a different temperature than exactly T$_{HO}$ (or more likely, a distribution of temperatures, since the strain relaxes over a finite lengthscale away from the parasitic regions). We resolve a sharp ($\sim 100$ mK) transition, and hence if there were two transitions, we should be able to see them.

	The fact that our results are consistent between sample S1 and S2, and we see only a single sharp transition at precisely T$_{HO}$, confirms that the elastic discontinuities are from the HO transition, and not due to strain-induced AFM.
		
	\newpage	
		
%% file: main_final.bbl
\begin{thebibliography}{10}

\bibitem{mydosh:rmp2011a}
J.~A. Mydosh, P.~M. Oppeneer, {\it Rev. Mod. Phys.\/} {\bf 83}, 1301 (2011).

\bibitem{mydosh:pm2014a}
J.~Mydosh, P.~Oppeneer, {\it Philosophical Magazine\/} {\bf 94}, 3642 (2014).

\bibitem{RamshawPNAS}
B.~J. Ramshaw, {\it et~al.\/}, {\it Proceedings of the National Academy of
  Sciences\/} {\bf 112}, 3285 (2015).

\bibitem{carrasquilla:np2017a}
J.~Carrasquilla, R.~G. Melko, {\it Nature Physics\/} {\bf 13}, 431 EP  (2017).

\bibitem{Ouyang:2018}
R.~Ouyang, S.~Curtarolo, E.~Ahmetcik, M.~Scheffler, L.~M. Ghiringhelli, {\it
  Phys. Rev. Materials\/} {\bf 2}, 083802 (2018).

\bibitem{Bohrdt:2018}
A.~Bohrdt, {\it et~al.\/}, {\it Nature Physics\/} {\bf 15}, 921 (2019).

\bibitem{Rem:2018}
B.~S. Rem, {\it et~al.\/}, {\it Nature Physics\/} {\bf 15}, 917 (2019).

\bibitem{Zhang:2018}
Y.~Zhang, {\it et~al.\/}, {\it Nature\/} {\bf 570}, 484 (2019).

\bibitem{Yamaji:2019}
Y.~Yamaji, T.~Yoshida, A.~Fujimori, M.~Imada, Hidden self-energies as origin of
  cuprate superconductivity revealed by machine learning (2019).

\bibitem{harima:jpsj2010a}
H.~Harima, K.~Miyake, J.~Flouquet, {\it Journal of the Physical Society of
  Japan\/} {\bf 79}, 033705 (2010).

\bibitem{ohkawa:jpcm1999a}
F.~J. Ohkawa, H.~Shimizu, {\it Journal of Physics: Condensed Matter\/} {\bf
  11}, L519 (1999).

\bibitem{santini:prl1994a}
P.~Santini, G.~Amoretti, {\it Phys. Rev. Lett.\/} {\bf 73}, 1027 (1994).

\bibitem{kiss:ap2004a}
A.~Kiss, P.~Fazekas, {\it Phys. Rev. B\/} {\bf 71}, 054415 (2005).

\bibitem{haule:np2009a}
K.~Haule, G.~Kotliar, {\it Nature Physics\/} {\bf 5}, 796 EP  (2009).

\bibitem{kusunose:jpsj2011a}
H.~Kusunose, H.~Harima, {\it Journal of the Physical Society of Japan\/} {\bf
  80}, 084702 (2011).

\bibitem{cricchio:prl2009a}
F.~Cricchio, F.~Bultmark, O.~Gr\aa{}n\"as, L.~Nordstr\"om, {\it Phys. Rev.
  Lett.\/} {\bf 103}, 107202 (2009).

\bibitem{Kung:2015}
H.-H. Kung, {\it et~al.\/}, {\it Science\/} {\bf 347}, 1339 (2015).

\bibitem{Kung:2016}
H.-H. Kung, {\it et~al.\/}, {\it Phys. Rev. Lett.\/} {\bf 117}, 227601 (2016).

\bibitem{kambe:pr2018a}
S.~Kambe, {\it et~al.\/}, {\it Phys. Rev. B\/} {\bf 97}, 235142 (2018).

\bibitem{thalmeier:pr2011a}
P.~Thalmeier, T.~Takimoto, {\it Phys. Rev. B\/} {\bf 83}, 165110 (2011).

\bibitem{hoshino:jpsj2013a}
S.~Hoshino, J.~Otsuki, Y.~Kuramoto, {\it Journal of the Physical Society of
  Japan\/} {\bf 82}, 044707 (2013).

\bibitem{rau:pr2012a}
J.~G. Rau, H.-Y. Kee, {\it Phys. Rev. B\/} {\bf 85}, 245112 (2012).

\bibitem{tonegawa:prl2012a}
S.~Tonegawa, {\it et~al.\/}, {\it Phys. Rev. Lett.\/} {\bf 109}, 036401 (2012).

\bibitem{fujimoto:prl2011a}
S.~Fujimoto, {\it Phys. Rev. Lett.\/} {\bf 106}, 196407 (2011).

\bibitem{ikeda:np2012}
H.~Ikeda, {\it et~al.\/}, {\it Nature Physics\/} {\bf 8}, 528 EP  (2012).

\bibitem{riggs:nc2015a}
S.~C. Riggs, {\it et~al.\/}, {\it Nature Communications\/} {\bf 6}, 6425 EP
  (2015).

\bibitem{note1}
It is also generally agreed that OP of the HO state orders at a finite
  wavevector of $Q=(0,0,1/2)$. Because our measurement occurs close to $Q=0$,
  i.e. at long wavelength, we are only concerned with the point-group symmetry
  of the OP, and not with its modulation in space.

\bibitem{okazaki:s2011a}
R.~Okazaki, {\it et~al.\/}, {\it Science\/} {\bf 331}, 439 (2011).

\bibitem{tonegawa2014direct}
S.~Tonegawa, {\it et~al.\/}, {\it Nature communications\/} {\bf 5}, 4188
  (2014).

\bibitem{Choi:2018}
J.~Choi, {\it et~al.\/}, {\it Phys. Rev. B\/} {\bf 98}, 241113 (2018).

\bibitem{Aji:2010}
V.~Aji, A.~Shekhter, C.~Varma, {\it Physical Review B\/} {\bf 81}, 064515
  (2010).

\bibitem{Rice:1995}
T.~Rice, M.~Sigrist, {\it Journal of Physics: Condensed Matter\/} {\bf 7}, L643
  (1995).

\bibitem{Read:2000}
N.~Read, D.~Green, {\it Phys. Rev. B\/} {\bf 61}, 10267 (2000).

\bibitem{Harrison:2019}
N.~Harrison, M.~Jaime, {\it arXiv preprint arXiv:1902.06588\/}  (2019).

\bibitem{Chandra2013}
P.~Chandra, P.~Coleman, R.~Flint, {\it Nature\/} {\bf 493}, 621 EP  (2013).
  Article.

\bibitem{Shekhter:2013}
A.~Shekhter, {\it et~al.\/}, {\it Nature\/} {\bf 498}, 75 (2013).

\bibitem{LuthiPRB}
B.~L\"uthi, T.~J. Moran, {\it Phys. Rev. B\/} {\bf 2}, 1211 (1970).

\bibitem{Fukase:1987}
T.~Fukase, {\it et~al.\/}, {\it Japanese Journal of Applied Physics\/} {\bf
  26}, 1249 (1987).

\bibitem{Bullock:1990}
G.~Bullock, B.~Shivaram, D.~G. Hinks, {\it Physica C: Superconductivity\/} {\bf
  169}, 497 (1990).

\bibitem{Wolf:1994}
B.~Wolf, {\it et~al.\/}, {\it Physica B: Condensed Matter\/} {\bf 199}, 167
  (1994).

\bibitem{Yanagisawa:2012}
T.~Yanagisawa, {\it et~al.\/}, {\it Journal of Physics: Conference Series\/}
  (IOP Publishing, 2012), vol. 391, p. 012079.

\bibitem{Yanagisawa:2018}
T.~Yanagisawa, {\it et~al.\/}, {\it Phys. Rev. B\/} {\bf 97}, 155137 (2018).

\bibitem{note2}
There are two $A_{1g}$ strains, $\epsilon_{xx}+\epsilon_{yy}$ and
  $\epsilon_{zz}$ with associated moduli $(c_{11}+c_{12})/2$ and $c_{33}$, as
  well as linear coupling between these two strains that produces a third
  modulus $c_{23}$. To simplify the notation we drop the sum over all three of
  these terms in the free energy.

\bibitem{Modic:2018b}
K.~A. Modic, {\it et~al.\/}, {\it Nature Communications\/} {\bf 9} (2018).

\bibitem{baumbachPhilMag}
R.~Baumbach, {\it et~al.\/}, {\it Philosophical Magazine\/} {\bf 94}, 3663
  (2014).

\bibitem{Cjaji:2001}
B.~C. Cs{\'a}ji, {\it Faculty of Sciences, Etvs Lornd University, Hungary\/}
  {\bf 24}, 48 (2001).

\bibitem{Buhot:2014}
J.~Buhot, {\it et~al.\/}, {\it Phys. Rev. Lett.\/} {\bf 113}, 266405 (2014).

\bibitem{IkedaStSol}
T.~Ikeda, K.~Fujibayashi, T.~Nagai, J.~Kobayashi, {\it physica status solidi
  (a)\/} {\bf 16}, 279 (1973).

\bibitem{GarlandJChemPhys}
C.~W. Garland, J.~S. Jones, {\it The Journal of Chemical Physics\/} {\bf 39},
  2874 (1963).

\bibitem{Matsuda:2011}
T.~D.~Matsuda, {\it et~al.\/}, {\it Journal of the Physical Society of Japan\/}
  {\bf 80}, 114710 (2011).

\bibitem{Gallagher:2016}
A.~Gallagher, {\it et~al.\/}, {\it Crystals\/} {\bf 6}, 128 (2016).

\bibitem{Okuda:2002}
N.~Okuda, T.~Suzuki, Z.~Mao, Y.~Maeno, T.~Fujita, {\it Journal of the Physical
  Society of Japan\/} {\bf 71}, 1134 (2002).

\bibitem{Greitemann:2019}
J.~Greitemann, K.~Liu, L.~Pollet, {\it Phys. Rev. B\/} {\bf 99}, 060404 (2019).

\bibitem{miglioriRSI}
A.~Migliori, J.~D. Maynard, {\it Review of Scientific Instruments\/} {\bf 76},
  121301 (2005).

\bibitem{Altmann:1994}
S.~Altmann, P.~Herzig, {\it Point-group theory tables\/} (Oxford, 1994).

\bibitem{RehwaldAIP1973}
W.~Rehwald, {\it Advances in Physics\/} {\bf 22}, 721 (1973).

\bibitem{MiyakePRL1986}
K.~Miyake, C.~M. Varma, {\it Phys. Rev. Lett.\/} {\bf 57}, 1627 (1986).

\bibitem{MatsudaPRL2001}
K.~Matsuda, Y.~Kohori, T.~Kohara, K.~Kuwahara, H.~Amitsuka, {\it Phys. Rev.
  Lett.\/} {\bf 87}, 087203 (2001).

\bibitem{Yanagisawa2014}
T.~Yanagisawa, {\it Philosophical Magazine\/} {\bf 94}, 3775 (2014).

\end{thebibliography}
